\documentclass[aps,preprint]{revtex4}
\usepackage[dvips]{graphics,graphicx}
\usepackage{amsmath}
\begin{document}

\title{Quantum transport in the flux rhombic lattice}
\author{P.~S.~Muraev$^2$}
\author{A.~R.~Kolovsky$^{1,2}$}
\affiliation{$^1$Kirensky Institute of Physics, 660036 Krasnoyarsk, Russia}
\affiliation{$^2$Siberian Federal University, 660041 Krasnoyarsk, Russia}
\date{\today}

\begin{abstract}
We analyse stationary current of the bosonic particles in the flux rhombic lattice connecting  two particle reservoirs. For vanishing inter-particle interactions the current is shown to monotonically decrease as the flux is increased and become strictly zero for the Peierls phase equal to $\pi$. Non-zero interactions modify this dependence and for moderate  interaction strength the current is found to be independent of the flux value. 
\end{abstract}
\maketitle

\section{Introduction}

Quantum transport through periodic structures is of permanent interest since the early days of Quantum mechanics \cite{Bloc28}. Recently this problem has been addressed for cold atoms in optical lattices where one of the research directions is the transport of Bose or Fermi atoms between two reservoirs which are connecting by a lattice \cite{Lebr18,Pros10,Znid10,Brud12,Ivan13,Kord15b,112,116}. Remarkably, under certain conditions this problem can be solved analytically that creates a reference point for studying different realistic systems.  Roughly, these conditions are the following: (i)  The reservoirs are Markovian so that one can justify a master equation for the reduced density matrix for the fermionic/bosonic carriers in the lattice; (ii) Inter-particle interactions are negligible so that we can use the formalism of the single-particle quantum mechanics; (ii)  The lattice has a simple structure and can be approximated by the linear tight-binding chain.  Violation of any of these conditions makes the system much harder for analysis but, simultaneously, introduces new effects.  In particular, it was shown in the recent work \cite{116} that inter-particle interactions change the ballistic transport regime, where the current is independent of the lattice length, into the diffusive transport regime, where the current is inverse proportional to the lattice length.

In the present work we extend the studies of Ref.~\cite{116} by considering the transport of Bose particles through the rhombic lattice (see Fig.~\ref{fig0} below). The Bloch spectrum of this lattice is known to consist of two dispersive bands and one flat band which is formed by the localised states. Moreover, by applying an external gauge field one can modify the dispersion relation of the dispersive bands, making them flat as well \cite{Vida00,Long14,Mukh15}. These relate the considered in the paper problem to the other fundamental problems like the role of flat bands in the quantum transport \cite{Khom16,109,Kim20} and the stability of the localised states in the presence of inter-particle interactions \cite{Vida00,Dani20,preprint}.

\section{The system}

The elementary cell of the rhombic lattice consists of three sites which we denote by letters $C_m$, $A_m$, and  $B_m$ where the subindex $m$ denotes the cell number, see Fig.~1. We assume the presence of the magnetic flux through a plaque characterised by the Peierls phase $\Phi$. Then the Bloch bands are given by the equation 
\begin{equation}
\label{1}
\epsilon_0(\kappa)=0 \;,\quad \epsilon_\pm(\kappa)=\pm J\sqrt{1+\cos(\Phi/2)\cos(\kappa-\Phi/2)} \;,  
\end{equation}
where $\kappa$ is the quasimomentum. Of a particular interest is the case $\Phi=\pi$ where the dispersive bands $\epsilon_\pm(\kappa)$ become flat, $\epsilon_\pm(\kappa)=\pm J$.
\begin{figure}[t]
\includegraphics[width=12.0cm,clip]{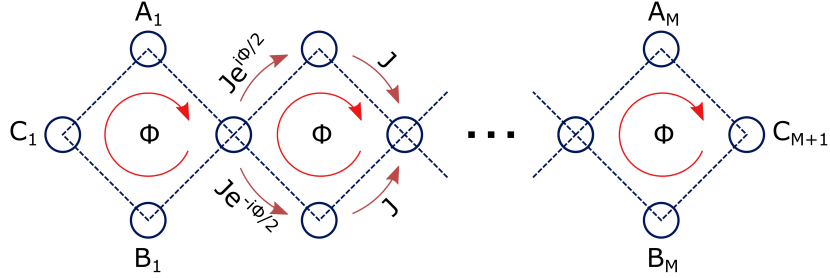}
\caption{Flux rhombic lattice consisting of $M=3$ rhombs. The flux is characterised by the Peierls phase $\Phi$ which determines the values of the hopping matrix elements.}
\label{fig0}
\end{figure} 

The system Hamiltonian is of the Bose-Hubbard type,
\begin{equation}
\label{2}
\widehat{H}= -\frac{1}{2}\sum_{l,m} \left(J_{l,m}  \hat{a}^\dag_{m}\hat{a}_l +h.c.\right)
  +\frac{U}{2} \hat{n}_l(\hat{n}_l-1)  \;,
 \end{equation}
where $|J_{l,m}|=J$. We chose to work with the gauge where the matrix element $J_{C,A}=Je^{i\Phi/2}$, the matrix element  $J_{C,B}=Je^{-i\Phi/2}$, and the remaining two matrix elements equal $J$.  For the numerical purpose we reorder the lattice sites as $C_1,A_1,B_1,C_2,A_2,B_2,\ldots$. (Thus, for example, the index $l=5$ corresponds to the $A$ site in the second cell.)  

We are interested in the transport of Bose particles across the rhombic lattices where particles come from the left reservoir into the site $C_1$ and go out of the lattice into the right reservoir from the site $C_{M+1}$. Thus, the master equation for the reduced density matrix ${\cal R}(t)$ of the carriers in the lattice has the form \cite{Ivan13,Kord15b,112,116}
\begin{equation}
\label{3}
\frac{d {\cal R}}{dt}=-i[\widehat{H},{\cal R}] + {\cal L}_{source}({\cal R}) +  {\cal L}_{drain}({\cal R}) \;,
\end{equation}
where the operator ${\cal L}_{source}({\cal R})$,
\begin{eqnarray}
\label{4}
 {\cal L}_{source}({\cal R})=-\frac{\gamma_{\rm L}}{2}\left[
 (\bar{n}_{\rm L}+1)(\hat{a}_1^\dagger \hat{a}_1{\cal R}-2\hat{a}_1{\cal R}\hat{a}_1^\dagger + {\cal R}\hat{a}_1^\dagger\hat{a}_1)
 + \bar{n}_{\rm L}(\hat{a}_1\hat{a}_1^\dagger{\cal R}-2\hat{a}_1^\dagger{\cal R}\hat{a}_1 + {\cal R}\hat{a}_1\hat{a}_1^\dagger) \right]   \;,
 \end{eqnarray}
and the operator ${\cal L}_{drain}({\cal R})$,
\begin{eqnarray}
\label{5}
 {\cal L}_{drain}({\cal R})=-\frac{\gamma_{\rm R}}{2}\left[
 (\bar{n}_{\rm R}+1)(\hat{a}_L^\dagger \hat{a}_L{\cal R}-2\hat{a}_L{\cal R}\hat{a}_L^\dagger + {\cal R}\hat{a}_L^\dagger\hat{a}_L)
 + \bar{n}_{\rm R}(\hat{a}_L\hat{a}_L^\dagger{\cal R}-2\hat{a}_L^\dagger{\cal R}\hat{a}_L + {\cal R}\hat{a}_L\hat{a}_L^\dagger) \right]   \;,
 \end{eqnarray}
take into account the coupling of the system to reservoirs with the mean particle density  $\bar{n}_{\rm L}$ (left reservoir) and  $\bar{n}_{\rm R}$ (right reservoir). Notice that the usage of the terms `source' and `drain' in Eqs.~(\ref{3}) implies that $\bar{n}_{\rm L}>\bar{n}_{\rm R}$.  

In the subsequent sections we solve Eq.~(\ref{3}) and calculate the single-particle density matrix (SPDM) of the carriers,
\begin{equation}
\label{6}
 \rho_{l,m}(t)={\rm Tr}[\hat{a}_l^\dagger\hat{a}_m {\cal R}(t)] \;,\quad 1\le l,m \le L \;, 
 \end{equation}
which suffices to predict the particle current between the reservoirs. The size of this matrix is obviously given by $L=3M+1$ where $M$ is the number of rhombs.

\section{Non-interacting particles $(U=0)$}

In the case of vanishing inter-particle interactions one can obtain a closed set of ordinary differential equations for the SPDM elements, 
\begin{eqnarray}
\label{7}
\frac{{\rm d}}{{\rm d} t} \rho_{l,m}=-i [\widehat{H},\hat{\rho} ]_{l,m} 
-\sum_{j=1,L}\frac{\gamma_j}{2}(\delta_{l,j}+\delta_{m,j})\rho_{l,m} 
+ \sum_{j=1,L} \bar{n}_j \delta_{l,j}\delta_{m,j} \;.
\end{eqnarray}
We mention that this equation is valid for any lattice as soon as the particles are injected in the first site of the lattice and withdrawn from the last site. In what follows,  to simplify equations, we assume  that the relaxation constants $\gamma_1\equiv \gamma_{\rm L}$ and  $\gamma_L\equiv\gamma_{\rm R}$ are the same and equal to $\gamma$.
\begin{figure}[b]
\includegraphics[width=8.0cm,clip]{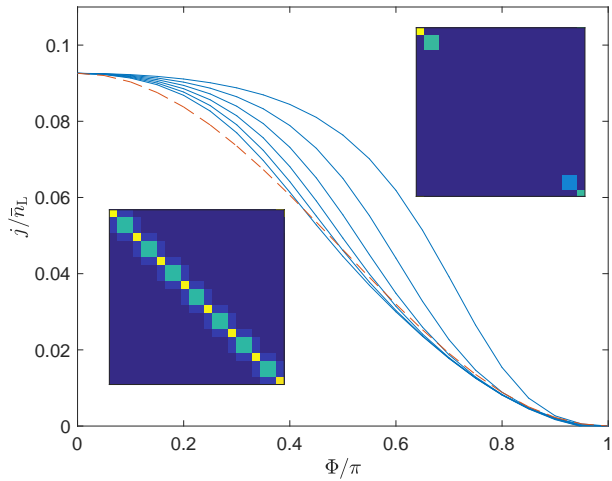}
\caption{The stationary current  as the function of $\Phi$ for $M=2,\ldots,7$.  Parameters are $\gamma_{\rm L}=\gamma_L=0.4$,  and $\bar{n}_{\rm R}/\bar{n}_{\rm L}=0.5$. The dashed red line is Eq.~(\ref{12}). The inserts show the elements of the stationary SPDM by the absolute values for $M=3$ and $\Phi=0$, lower-left conner, and $\Phi=\pi$, upper-right conner.}
\label{fig2}
\end{figure}

Let us first discuss the case $\Phi=0$.  The stationary SPDM of the bosonic carriers in the rhombic lattice is exemplified in the lower-left conner in Fig.~\ref{fig2} for $M=3$. Here the stationary populations of the $C$ sites are the same (except the first and last sites) and are given by the equation,
\begin{equation}
\label{8}
\rho_{l,l}^{(C)}= \frac{\bar{n}_{\rm L}+\bar{n}_{\rm R}}{2} \;,
\end{equation}
and populations of the $A$ and $B$ sites are one half of this value. The stationary current is
\begin{equation}
\label{9}
\bar{j}=j_0\frac{J\gamma}{J^2+\gamma^2/2} \frac{\bar{n}_{\rm L}-\bar{n}_{\rm R}}{2} \;.
\end{equation}
Comparing this equation with Eq.(34) in Ref.~\cite{116} ) we conclude that for $\Phi=0$ the rhombic lattice behaves similar to the simple linear lattice with the even sites given by the `sum' of the $A$ and $B$ sites of the rhombic lattice.

Next we analyse the case  $\Phi=\pi$, see the upper-right conner  in Fig.~\ref{fig2}.  As expected, here the propagation of particles across the lattice  is blocked by the destructive interference and and the current is strictly zero. For populations of the edge sites we have
\begin{equation}
\label{10}
\rho_{1,1}=\bar{n}_{\rm L} \;, \quad  \rho_{L,L}=\bar{n}_{\rm R} \;,
\end{equation}
and populations of the neighbouring $A$ and $B$ sites are one half of these values.  We also mention that the $A-B$ dimers at the lattice edges are in the antisymmetric (left edge) and the symmetric (right edge) states, i.e.,
\begin{equation}
\label{11}
\rho_{A,B}=\mp \sqrt{\rho_{A,A}\rho_{B,B}} \;.
\end{equation}

Unfortunately, there is no simple analytical expression for the stationary current for an arbitrary $\Phi$.  Moreover, the result depends on the lattice length.  The main panel  in Fig.~\ref{fig2} shows the stationary current in the system as the function of $\Phi$ for $2\le M \le 7$.  (Here we normalise the current to the particle density in the left reservoir.) It is seen that the curves rapidly converge to some limiting curve which can be approximated by the  relation
\begin{equation}
\label{12}
\bar{j}(\Phi) \approx \bar{j}(\Phi=0) \cos^2(\Phi/2) \;.
\end{equation}
One may naively assume that the dependence (\ref{12}) is given by the mean squared group velocity of a quantum particle in the lattice \cite{112}, however, this appears to be not the case. A justification of Eq.~(\ref{12}) remains the open problem.

\section{Interacting particles}

To treat the case of interacting particles we use the pseudo-classical approach (also known as the truncated Wigner function or truncated Husimi function approximations) which  was proved to be very accurate when analysing the current in the simple linear lattice \cite{116}. This approach reduces the master equation (\ref{3}) to the Fokker-Planck equation on the classical distribution function $f=f({\bf a},{\bf a}^*;t)$ defined in the multi-dimensional phase-space ${\bf a}=a_1,\ldots, a_L$,
\begin{equation}
\label{c1}
\frac{\partial f}{\partial t}=\{H,f\} + \sum_{l={\rm L},{\rm R}} \left[ {\cal G}^{(l)}(f) + {\cal D}^{(l)}(f) \right]  \;.
\end{equation}
In Eq.~(\ref{c1})  $H$ is the classical counterpart of the Bose-Hubbard Hamiltonian  (\ref{2}), $\{\ldots,\ldots\}$ denotes the Poisson brackets, and the terms in the square brackets are the Weyl images of the Lindblad operators  ${\cal L}_{source}({\cal R})$ and  ${\cal L}_{drain}({\cal R})$.  To clarify the mathematical structure of the  equation here we explicitly decompose these images into the friction terms,
\begin{equation}
\label{c2}
{\cal G}^{(l)}(f)=\frac{\gamma_l}{2}\left( a_l\frac{\partial f}{\partial a_l} + 2f+ a_l^*\frac{\partial f}{\partial a_l^*} \right) \;,
\end{equation}
and the diffusion terms  
\begin{equation}
\label{c0b}
{\cal D}^{(l)}(f)= D_l\frac{\partial^2 f}{\partial a_l \partial a_l^*}  \;,
\end{equation}
where the diffusion coefficients $D_{\rm L}$ and $D_{\rm R}$ are proportional to the reservoir particle densities $\bar{n}_{\rm L}$ and $\bar{n}_{\rm R}$, respectively.

Knowing the distribution function $f=f({\bf a},{\bf a}^*;t)$ the SPDM elements are calculated by taking the multi-dimensional integral
\begin{equation}
\label{13}
\rho_{l,m}(t)= \int a_l^* a_m  f({\bf a},{\bf a}^*;t) {\rm d} {\bf a} {\rm d} {\bf a}^* \;.
\end{equation}
We mention that the method is exact for $U=0$ and in the formal limit $U\rightarrow 0$, $\bar{n}_{\rm L}\rightarrow\infty$, $g=U\bar{n}_{\rm L}=const$. The main advantage of the approach is that, when we cannot solve Fokker-Planck equation (\ref{c1}) analytically, it is always possible to estimate $\rho_{l,m}(t)$ by mapping  this equation into the Langeven equation and then employing the Monte-Carlo simulation.
\begin{figure}[t]
\includegraphics[width=8.0cm,clip]{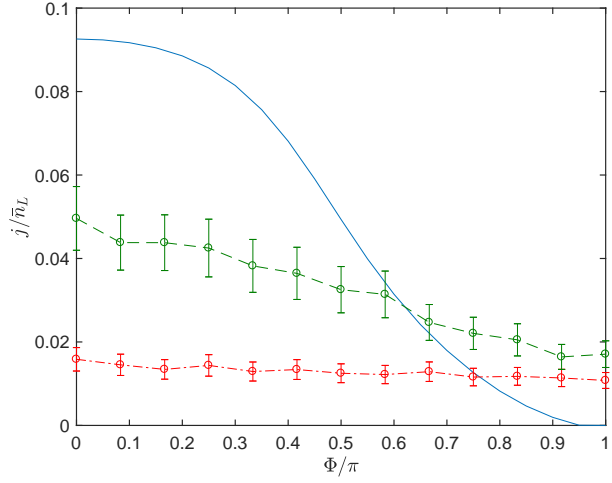}
\caption{The stationary current  as the function of $\Phi$ for $M=5$ and $g=0$, solid line, $g=0.7$, dashed line, and $g=2$, dash-dotted line.  The other parameters are the same as in the previous figure. The error bar indicates statistical error due to the Monte-Carlo simulation.}
\label{fig3}
\end{figure} 
\begin{figure}
\includegraphics[width=9.0cm,clip]{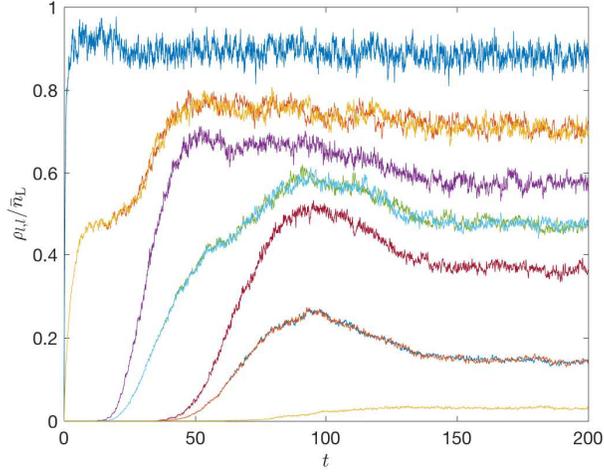}
\caption{Population of the lattice sites (i.e., diagonal elements of SPDM) as the function of time for $\Phi=\pi$ and $M=3$.   The other parameters are $\gamma_{\rm L}=\gamma_L=0.4$,  $\bar{n}_{\rm R}=0$, and $g=4$.  Average over 400 realizations. Different curves refer to  $\rho_{C,C}^{(1)}$, $\rho_{A,A}^{(1)}$  and $\rho_{B,B}^{(1)}$, $\rho_{C,C}^{(2)}$, $\rho_{A,A}^{(2)}$  and $\rho_{B,B}^{(2)}$,  etc., from top to bottom.}
\label{fig4}
\end{figure}

Fig.~\ref{fig3} compares the stationary current in the flux rhombic lattice $g=0$, $g=0.7$, and $g=2$.  It is seen that for $\Phi$ close to zero interactions suppress the current. This is consistent with the results of Ref.~\cite{116} and, in fact, is an indication of the transition from the ballistic transport regime to the diffusive regime. For $\Phi$ close to $\pi$, however, the current is  enhanced and has a finite value even for $\Phi=\pi$ where transport is forbidden due to the localisation effect if $g=0$ . Thus, interactions destroy the localisation. This is consistent with results of Ref.~\cite{Vida00}, which analyses the energy spectrum of two interacting fermions in the flux rhombic lattice, and results of Ref.~\cite{preprint}, which specifically addresses the stability of the localised states against interactions. In particular, it was shown in the latter work that the antisymmetric (or symmetric, depending on the chosen gauge) localised $A-B$ state is subject to dynamical instability which leads to excitation of the unprotected symmetric (antisymmetric) $A-B$ state. Because the developing of instability takes some time, in the rhombic lattice we have very long transient regime for $\Phi=\pi$. During this transient we observe subsequent populations of the $C$ sites with the time delay given by the instability time, see Fig.~\ref{fig4}. When all $C$ sites are populated, the system reaches the steady-state regime with the diffusive-like transport from the left to the right reservoirs. 

\section{Conclusion}

We studied the current of non-interacting and interacting bosonic carriers across the flux rhombic lattice. In the case of vanishing inter-particle interactions the transport is ballistic and the current is determined by the interference effects due to the presence of two alternative passes between the $C$ sites. For zero flux the interference is constructive and the current is maximal. On the contrary, for the flux corresponding to the Peierls phase $\Phi=\pi$ the interference is completely destructive and the current is zero. For the intermediate value of $\Phi$ the current was found to be approximately proportional to $\cos^2(\Phi/2)$. 

Unlike for vanishing interactions, for moderate interaction strength $g\sim J$ the current is mainly determined by the interaction effects which change the ballistic transport regime into diffusive transport. It can be  expected from general arguments that diffusion destroys interference. This expectation was fully confirmed by the straightforward numerical analysis of the system dynamics where the stationary current of the bosonic carries was found to be essentially independent of the flux.

This work has been supported through Russian Science Foundation grant N19-12-00167.



\begin{thebibliography}{99}

 \bibitem{Bloc28}
 F. Bloch,
 {\em \"Uber die Quantenmechanik der Elektronen in Kristallgittern},
 Zeitschrift f\"ur Physik {\bf 52}, 555 (1928).

\bibitem{Lebr18}
M. Lebrat, P. Grisins, D. Husmann, S. H\"ausler, L. Corman, T. Giamarchi, J.-Ph. Brantut, and T. Esslinger, 
{\em Band and Correlated Insulators of Cold Fermions in a Mesoscopic Lattice}, 
Phys. Rev. X {\bf 8}, 011053 (2018).

\bibitem{Pros10}
T. Prosen, B. \^Zunkovi\^c,   
{\em Exact solution of Markovian master equations for quadratic Fermi systems: thermal baths, open XY spin chains and non-equilibrium phase transition},
New J.  of Phys. {\bf 12}, 025016 (2010).
%
\bibitem{Znid10}
M. Znidaric,
{\em Exact solution for a diffusive nonequilibrium steady state of an open quantum chain},
J. Stat. Mech. 2010, L05002 (2010).

\bibitem{Brud12}
M. Bruderer, W. Belzig, {\em Mesoscopic transport of fermions through an engineered optical lattice connecting two reservoirs}, 
Phys. Rev. A {\bf 85}, 013623 (2012); 

\bibitem{Ivan13}
A. Ivanov, G. Kordas, A. Komnik, and S. Wimberger, 
{\em Bosonic transport through a chain of quantum dots}, 
Eur. Phys. J. B {\bf 86}, 345 (2013).


\bibitem{Kord15b}
G. Kordas, D. Witthaut, S. Wimberger, 
{\em Non-equilibrium dynamics in dissipative Bose-Hubbard chains}, 
Ann. Phys. (Berlin) {\bf 527}, 619 (2015).

\bibitem{112}
A. R. Kolovsky, Z. Denis, and S. Wimberger,
{\em Landauer-B\"uttiker equation for bosonic carriers},
Phys. Rev. A {\bf 98}, 043623 (2018).

\bibitem{116}
A. A. Bychek, P. S. Muraev, D. N. Maksimov, and A. R. Kolovsky,
{\em Open Bose-Hubbard chain: Pseudoclassical approach},
Phys. Rev. E {\bf 101}, 012208 (2020).




\bibitem{Vida00}
J. Vidal, B. Doucot, R. Mosseri, and P. Butaud, 
{\em Interaction induced delocalisation for two particles in a periodic potential}, 
Rev. Lett. {\bf 85}, 3906 (2000).

\bibitem{Long14}
S. Longhi,
{\em Effective magnetic fields for photons in waveguide and coupled resonator lattices},
Opt. Lett. {\bf 38}, 3570 (2013). 


\bibitem{Mukh15}
S.~Mukherjee and R.~R.~Thomson,
{\em Observation of localized flat-band modes in a quasi-one-dimensional photonic rhombic lattice},
Opt. Lett. {\bf 40}, 5443 (2015).




\bibitem{Khom16}
R. Khomeriki, S. Flach,
{\em Landau-Zener Bloch oscillations with perturbed flat bands},
Phys. Rev. Lett. {\bf 116}, 245301 (2016).

\bibitem{109}
A. R. Kolovsky,  A. Ramachandran, and S. Flach,
{\em Topological flat Wannier-Stark bands},
Phys. Rev. B {\bf 97}, 045120 (2018).


\bibitem{Kim20}
Kun Woo Kim, A. Andreanov, S. Flach,
{\em Anomalous transport in a topological Wannier-Stark ladder}, 
Phys. Rev. Research {\bf 2}, 023067 (2020).

\bibitem{Dani20}
C. Danieli, A. Andreanov, S. Flach, 
{\em Many-Body Flatband Localization},
arXiv:2004.11928 (2020).

\bibitem{preprint}
D. N. Maksimov, A. A. Bychek,  and A. R. Kolovsky,
{\em Decay of symmetry protected quantum localised states},
arXiv:2005.14432 (2020).


\end{thebibliography}
\end{document}